\documentclass[a4paper,twoside,reqno]{bjp}
\usepackage{graphicx}
\usepackage{cite}
\usepackage{amssymb,amsmath,amscd,amsthm}
\usepackage{times}
\usepackage{pbox}

\usepackage[bookmarks=false]{hyperref}
\hypersetup{%
    colorlinks=true,        % false: boxed links; true: colored links
    linkcolor=blue,          % color of internal links (change box color with linkbordercolor)
    citecolor=blue,         % color of links to bibliography
    urlcolor=blue           % color of external links
    }

\usepackage{geometry}
\allowdisplaybreaks

\begin{document}

\title{Global performance of covariant density functional theory in description of charge radii and related indicators.}
% Insert the title

\runningheads{Afanasjev, Perera and Ring}{Global performance of CDFT in description of charge radii and related indicators}
% In case the authors are more then three, put the name of the first author followed by the Latin `et al.'

\begin{start}{%
\author{A.V.Afanasjev}{1},
% The second argument connects author(s) with addresses
\author{U.C.Perera}{1},
% The second argument connects author(s) with addresses
\author{P. Ring}{2}
% The second argument connects author(s) with addresses

\address{Department of Physics and Astronomy, Mississippi
	State University, MS 39762}{1}
\address{Fakult\"at f\"ur Physik, Technische Universit\"at M\"unchen,
	D-85748 Garching, Germany}{2}

\received{Day Month Year (Insert date of submission)}
% Insert date of submission
}

\begin{Abstract}
 A short review of existing efforts to understand charge radii  and related indicators on a
global scale within the covariant density functional theory (CDFT) is presented.
Using major classes of covariant energy density functionals (CEDFs), the  global accuracy
of the description of experimental absolute and differential charge radii within the CDFT
framework has been established. This assessment is supplemented by an evaluation of
theoretical statistical and systematic uncertainties in the description of charge radii.  New results
on the accuracy of the  description of differential charge radii in deformed actinides and light
superheavy nuclei are presented and the role of octupole deformation in their reproduction
is evaluated. Novel mechanisms leading to odd-even staggering in charge radii are
discussed. Finally, we analyze the role of self-consistency effects in an accurate description of
differential charge radii.
\end{Abstract}

\begin{KEY}
Covariant density functional theory, charge radii, differential charge radii,
theoretical uncertainties, self-consistency effects
\end{KEY}
\end{start}

%%%%%%%%%%%%%
\section{Introduction}
%%%%%%%%%%%%%

  The charge radii are among the most fundamental properties of atomic nuclei.
Essential information on the saturation density of symmetric nuclear matter
is imprinted  into them.  They also depend on the properties of nuclear forces
and nuclear many-body dynamics. Charge radii have been in the focus of
experimental  and theoretical studies for a long time and, in the last decade,
there was a significant increase in the interest to them due to the availability of
new facilities and new techniques. Absolute charge radii are determined
via the studies of muonic spectra and electronic scattering experiments
\cite{AM.13,CMP.16}.  However, such experiments are impossible in radioactive
elements.   Laser spectroscopy is a more flexible tool in that respect and  allows a
significantly higher precision of the measurements of the changes of  charge radii
within the isotopic chain \cite{AM.13,CMP.16}.

 The CDFT \cite{VALR.05} is one of the modern theoretical tools which has
been  used with reasonable success to the description of charge radii and related
indicators.  For  example, the first-ever successful description of the kink in charge radii of the lead
isotopes has been achieved in the CDFT framework in Ref. \cite{SLR.93} and in the
90ies it followed by a study of differential charge radii in spherical even-even
nuclei of the Ca, Sn, and Pb isotopic chains in Ref.\ \cite{RF.95}. However, 
only during the last seven years did new CDFT results appear, dedicated to the study of 
charge radii, related indicators, and other connected issues.
The goal of this paper is to present a short review of different facets of physics of charge radii
in CDFT and to provide new clarifying results. All the quoted and new calculations
are  performed within the framework of relativistic Hartree-Bogolibov (RHB)
approach with separable pairing interaction of Ref.\ \cite{TMR.09}. The technical
details of the RHB approach and numerical details of the calculations are presented in 
Ref.\ \cite{AARR.14}. The calculations are carried out with the CEDFs (such 
as NL3* \cite{NL3*}, DD-ME2 \cite{DD-ME2}, DD-ME$\delta$  \cite{DD-MEdelta}, DD-PC1 
\cite{DD-PC1}, PC-PK1 \cite{PC-PK1}  and NL5(C), NL5(D) and NL5(E) \cite{AAT.19}) representing 
the major classes of the CDFT models.

%%%%%%%%%%%%%%%%%
\section{Physical observables}
%%%%%%%%%%%%%%%%%

  We focus here on physical observables,
which provide information on charge radii and related indicators. The charge radii were
calculated from the corresponding point proton radii as (see Ref.\ \cite{AARR.14})
\begin{equation}
r_{ch} = \sqrt{<r^2>_p + 0.64}\,\,\,\, {\rm fm}
\label{r_charge}
\end{equation}
where $<r^2>_p$ stands for proton mean-square point radius
and the factor 0.64 accounts for the finite-size effects of the
proton.

    In addition, two differential indicators are commonly used  to facilitate the
quantitative comparison of the experimental results with those from theoretical
calculations. One of them is the differential mean-square charge radius
\begin{eqnarray}
\delta \left < r^2 \right>_p^{N,N'}
                                                     =r^2_{ch}(N) - r^2_{ch}(N')
\label{diff-radii}
\end{eqnarray}
where $N'$ is the neutron number of the reference nucleus. Another is the
three-point indicator
\begin{eqnarray}
\Delta \langle r^2 \rangle^{(3)}(N)
= \frac{1}{2} \left[ r_{ch}^2(N-1) + r_{ch}^2(N+1) - 2  r_{ch}^2(N) \right]
\label{Delta-3}
\end{eqnarray}
which quantifies OES in charge radii.

%%%%%%%%%%%%%%%%%%%%%%%%%%%%%%%%%%%%%%%
\section{Global performance of the CDFT in the description of charge radii}
%%%%%%%%%%%%%%%%%%%%%%%%%%%%%%%%%%%%%%%

%%%%%%%%%%%%%%%%%%%%%%%%%%%%%%%%%%%%%%%
\begin{figure}[htb]
\centerline{\includegraphics[angle=-90,width=11.0cm]{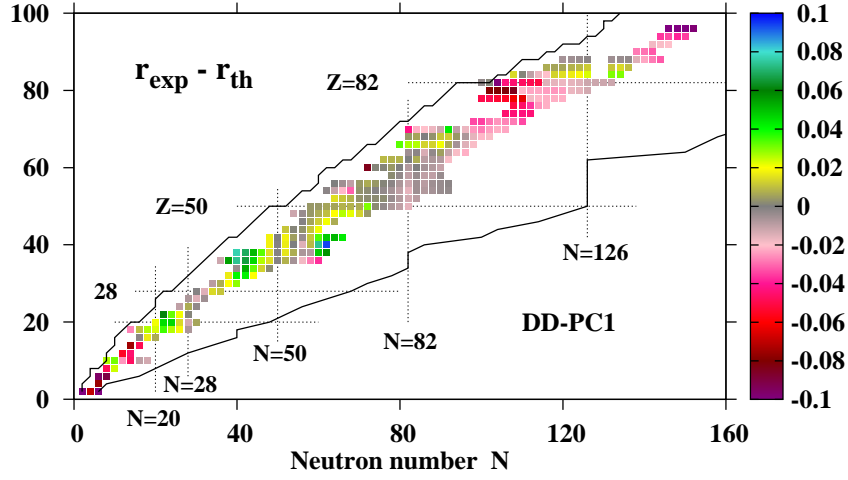}}
\caption{The difference between measured and calculated charge
radii  $r_{ch}$ for the CEDF DD-PC1.  Two-proton and two-neutron drip lines
obtained with this functional are shown by solid black lines. The experimental
data are taken from Ref.\ \cite{AM.13}. Based on figure from Ref.\ \cite{AA.16}.
}
\label{r_ch_dif}
\end{figure}
%%%%%%%%%%%%%%%%%%%%%%%%%%%%%%%%%No-region

The global performance of CDFT in the description of charge radii for major
classes of CEDFs has been studied in Refs.\ \cite{AARR.14,AA.16,AAT.19}, and
Fig.\ \ref{r_ch_dif} presents an example of a comparison between theory and
experiment.  The following rms deviations $\Delta r_{ch}^{rms}$ between
calculated and experimental charge radii have been reported:
$\Delta r_{ch}^{rms}=0.0283$ fm  for CEDF NL3*,
$\Delta r_{ch}^{rms}=0.0230$ fm  for DD-ME2,
$\Delta r_{ch}^{rms}=0.0329$ fm  for DD-ME$\delta$*,
$\Delta r_{ch}^{rms}=0.0253$ fm  for DD-PC1,
$\Delta r_{ch}^{rms}=0.0284$ fm  for NL5(C),
$\Delta r_{ch}^{rms}=0.0277$ fm  for NL5(D),  and
$\Delta r_{ch}^{rms}=0.0288$ fm  for NL5(E).
Note that the DD-PC1 functional has been adjusted only to nuclear
binding energies while other functionals include the information on
charge radii in their fitting protocols. The average rms deviations
of calculated $r_{ch}$ from experimental ones for this group of
functionals are at the level of 0.028 fm. Considering that the
average experimental rms charge radius in the nuclear
chart is around 4.8 fm (see, for example, Fig. 23 in Ref.
\cite{AARR.14} and Figs. 2-4 in Ref.\ \cite{AM.13}), this amounts to  a high
average precision of 0.58\% in the prediction of charge radii.

 Fig.\ \ref{r_ch_dif} reveals several regions on the nuclear chart
which  exhibit enhanced (with respect of average trend) differences
$(r_{ch}^{exp} - r_{ch}^{th})$  between experimental and calculated
charge radii.  These are (i) proton-rich nuclei around Hg ($Z=80$),
(ii) proton-rich nuclei around Sr ($Z=38$) located below the $N=50$ shell
closure and (iii) light and very light nuclei in the region with $Z<20$. The
detailed analysis of the nuclei in the regions (i) and (ii) performed
in Refs.\ \cite{UAR.21,FP.75, DGLGHPPB.10,Otten,Hg-mid} clearly indicates
that shape coexistence and the correlations beyond mean-field are the sources 
of these discrepancies.  Note that the latter are
neglected in our studies.  The importance of the correlations beyond mean-field 
is also expected to be increased in light and very light nuclei, the potential
energy surfaces of which are very frequently soft in collective coordinates.
The fact that the accuracy of the description of the masses declines with
decreasing mass also points in that direction (see discussion in Sec. V of
Ref.\ \cite{AARR.14}). Thus, the inclusion of the correlations beyond mean-field 
in the regions (i-iii) may improve the description of charge radii and
binding energies. The accuracy of the description of charge radii also deteriorates in
the actinides (see Fig.\ \ref{r_ch_dif}). However, this is most likely due to the
extrapolations used in Ref.\ \cite{AM.13} for the definition of absolute charge
radii since they have not been measured experimentally in the $Z>83$
nuclei (uranium is an exception).

%%%%%%%%%%%%%%%%%%%%%%%%%%%%%%%%%%%%%%%%%
\section{Statistical and systematic uncertainties in the description of charge radii}
%%%%%%%%%%%%%%%%%%%%%%%%%%%%%%%%%%%%%%%%%

Because of underlying physical approximations in modeling of the nuclear many-body problem it is
important to estimate statistical and systematic theoretical uncertainties in the predictions
of charge radii. This is especially important when one  deals with the extrapolations beyond
the known regions, as, for example, in particle number or deformation, since experimental
data which act as a substitute of an exact solution are not available there. Such estimates
are also required for the evaluation of the predictive power of the models and the robustness of
their predictions.

\begin{figure}[htb]
\centerline{\includegraphics[angle=0,width=12.0cm]{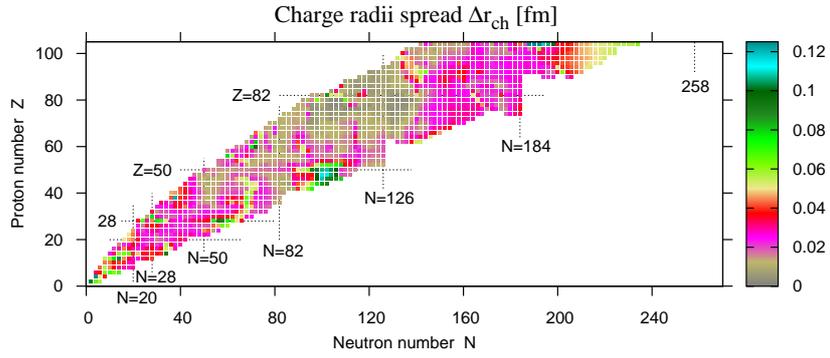}}
\caption{Systematic uncertainties in the description of charge radii. From Ref.\
\cite{AARR.14}.}
\label{system-errors}
\end{figure}

 Systematic theoretical uncertainties emerge from underlying theoretical approximations
about the form of the energy density functional. In the DFT framework, there are two major sources
of these  approximations, namely, the range of interaction and the form of the density
dependence  of the effective interaction \cite{BHR.03,BB.77,VALR.05}. In reality, only the
lower limit of these uncertainties is possible to estimate (see Refs.\  \cite{AARR.14,DNR.14,AAT.19})
and even then, it contains a degree of arbitrariness related to the choice of the set of the functionals.
In our works, it is estimated via the spread of theoretical predictions within a set of selected
functionals (NL3*, DD-ME2, DD-PC1, DD-ME$\delta$) as
\begin{equation}
\Delta r_{ch}(Z,N) = | r_{ch}^{max} (Z,N) -  r_{ch}^{min} (Z,N) |
\end{equation}
where $r_{ch}^{max} (Z,N)$ and $r_{ch}^{min} (Z,N)$ are the largest and smallest values
of the charge radii obtained with the four employed CEDFs for the $(Z,N)$ nucleus
\cite{AARR.14}. The
charge radii spreads are shown in Fig.\ \ref{system-errors}.  For the nuclei shown in Fig.\
\ref{r_ch_dif} they are on the level of 0.01 fm for
the majority of medium mass and heavy nuclei but  become larger for $Z\leq 40$
nuclei where they exceed 0.02 fm in most of the nuclei.  The latter feature is most likely due
to softer potential energy surfaces of such nuclei and thus stronger dependence of the results
on the details of the underlying single-particle structure. As a result, in some nuclei the systematic
theoretical uncertainties are either  comparable or larger than the differences between
experimental and calculated charge radii (compare Figs. \ref{system-errors} and
\ref{r_ch_dif}). This fact has to be kept in mind when performing a detailed comparison
between theory and experiment.

%%%%%%%%%%%%%%%%%%%%%%%%%%%
%%%%%%%%%%%%%%%%%%%%%%%%%%%%%%%%%No-region

%%%%%%%%%%%%%%%%%%%%%%%%%%%
\begin{figure}[htb]
\centerline{\includegraphics[width=7.0cm]{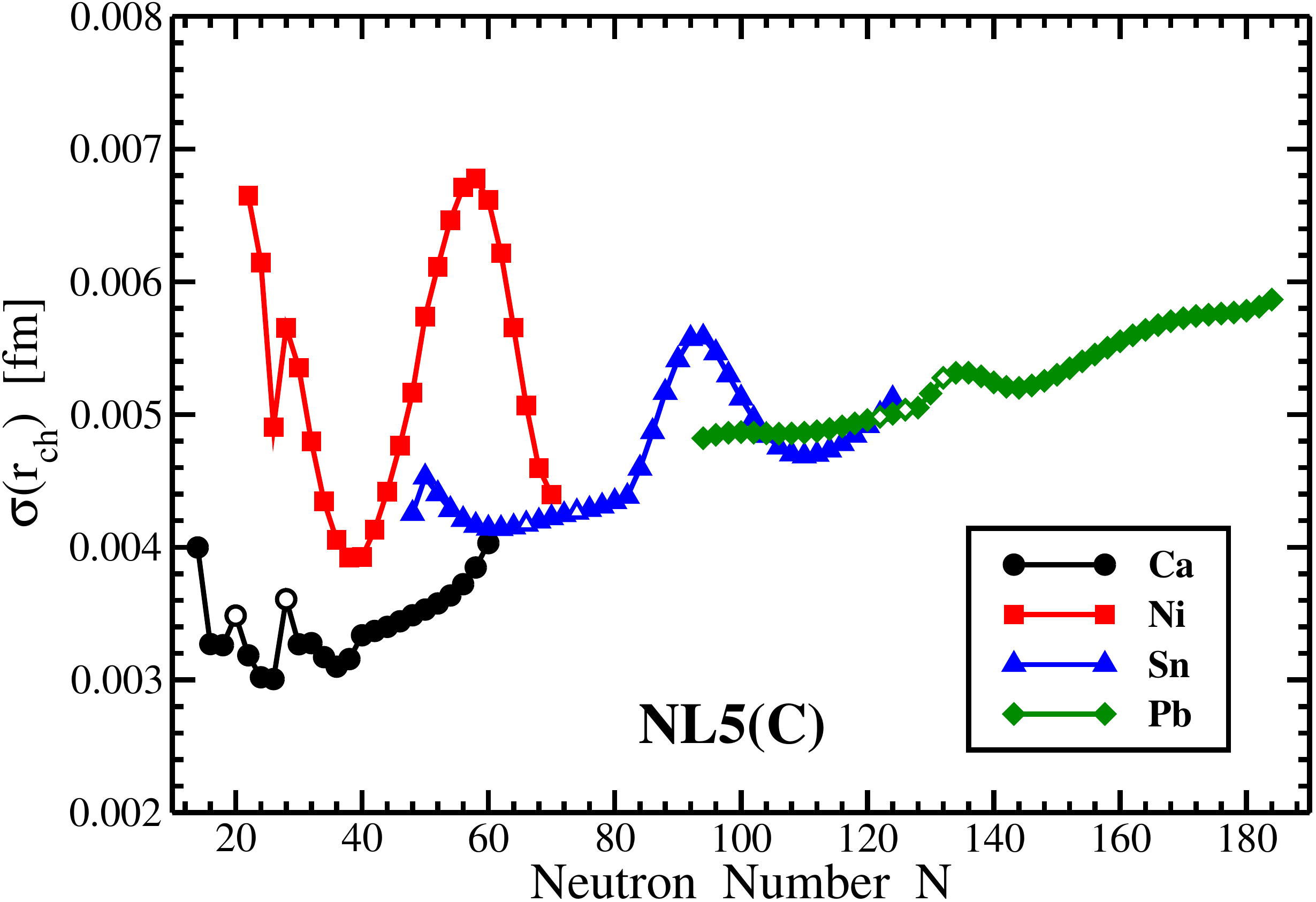}}
\caption{  The propagation of statistical errors in charge radii with neutron number for
the Ca (Z = 20), Ni (Z = 28), Sn (Z = 50),  and Pb (Z = 82) isotopic chains.  The RHB
calculations have been performed at spherical shape with the NL5(C) functional  for
all even-even nuclei located between the two-proton and two-neutron drip lines (see
Refs.\ \cite{AAT.19} for details). Open symbols are used to indicate the nuclei whose
experimental data have been used in the fitting protocol of this functional. From Ref.\
\cite{AAT.19}.}
\label{stat-errors}
\end{figure}
%%%%%%%%%%%%%%%%%%%%%%%%%%%%%%%%%No-region

 Statistical errors $\sigma(r_{ch})$, which apply only to a given functional, are due to
the details of the fitting protocol, such as the choice of experimental data and the selection of
adopted errors (see Refs.\ \cite{DNR.14,AAT.19}). They are shown in Fig.\ \ref{stat-errors} for
the NL5(C) functional for several isotopic chains of spherical nuclei. However, similar magnitude
of $\sigma(r_{ch})$ is expected both for other functionals and for deformed nuclei \cite{AAT.19}.
One can see that, on average, they are by a factor of approximately 6  smaller than global rms
deviations between theory and experiment. They are also substantially smaller than systematic
theoretical uncertainties for charge radii.  Note that the  $\sigma(r_{ch})$ values show some fluctuations as a
function of neutron number due  to the underlying shell structure; they become especially
pronounced in the Ni isotopes.  These fluctuations are expected to wash out in deformed nuclei
because of  a  more equal distribution of deformed single-particle states originating from different spherical
subshells.  On average, the $\sigma(r_{ch})$ values shown in Fig.\ \ref{stat-errors}  display a very
modest increase on going from the two-proton to the two-neutron drip line.  This is in contrast to the
results obtained in the Skyrme DFT calculations with the UNEDF0 functional (see Ref.\ \cite{GDKTT.13})
which for the nuclei near the proton drip line shows $\sigma(r_{ch})$ values comparable to those
obtained in the CDFT(NL5(C)) calculations, but are by a factor of 17-33 larger at the neutron drip line
(see Ref.\ \cite{AAT.19}). Based on these considerations, one can conclude that statistical errors
represent only a relatively small part of the total theoretical error in  the prediction of charge radii and
in many cases they can be neglected.

%%%%%%%%%%%%%%%%%%%%%%%%%%%%%%%%%%%%%%%%%
\section{Differential charge radii: the examples of the Sn/Gd region and the region of
actinides and light superheavy nuclei}
%%%%%%%%%%%%%%%%%%%%%%%%%%%%%%%%%%%%%%%%%

\begin{figure}[htb]
\centerline{\includegraphics[width=1.0\textwidth]{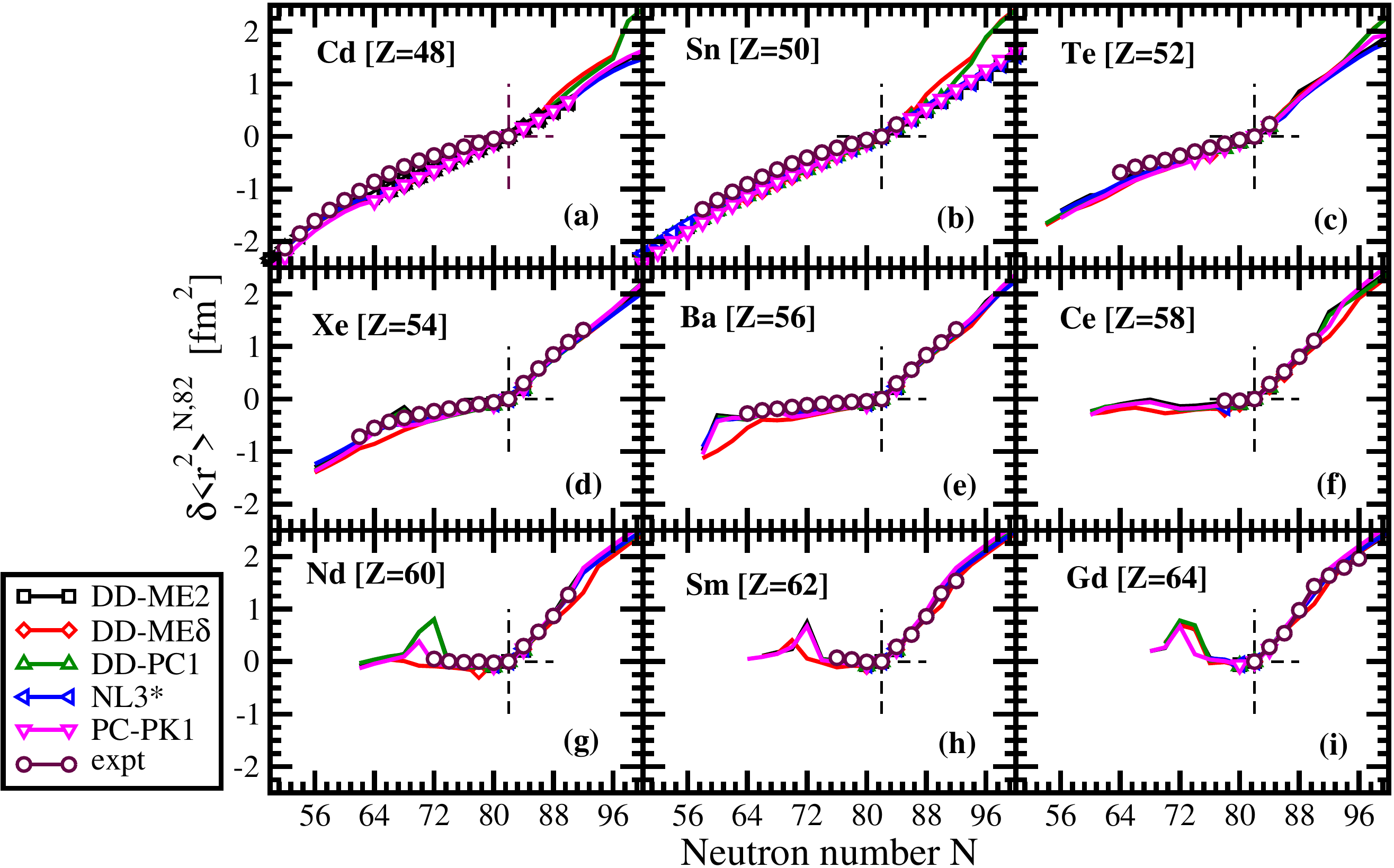}}
\caption{The experimental and calculated differential charge radii $\delta \left<r^2 \right>^{N,N'}$
of the indicated isotopic chains. The experimental data are taken from Ref.\ \cite{AM.13}. The calculations
are performed with the indicated CEDFs. Theoretical results are shown by lines with symbols and by
only lines for spherical and deformed ground states, respectively.
}
\label{SN/Gd-region}
\end{figure}

The differential charge radius $\delta \left<r^2 \right>^{N,N'}$  (see Eq.(\ref{diff-radii})) is an indicator
which is very sensitive to the shell closures,  underlying single-particle structure,
pairing  interaction, and occupation pattern of the single-particle states and their
evolution with neutron number (see Refs.\ \cite{UAR.21,OES-Pb-Hg.21-PRC} and
references quoted therein).  Their systematic global investigation has been performed within
the CDFT framework in Ref.\ \cite{UAR.21} and theoretical results were compared
with experimental data which crosses the neutron shell closures at N = 28, 50, 82, and 126. 
An example of such a comparison is presented in Fig.\ \ref{SN/Gd-region}.
Absolute differential radii of different isotopic chains and their relative properties are
reasonably well described in model calculations in the cases when the mean-field
approximation is justified.  The observed clusterization of differential charge radii of
different isotopic chains (see Ref.\ \cite{GV.20}) is well described above the $N = 50$ and
$N = 126$ shell closures (see Ref.\ \cite{UAR.21}). However, a similar clusterization
above the $N = 28$ and $N = 82$ shell closures is not that well reproduced due to
deficiencies in the underlying single-particle structure.

%%%%%%%%%%%%%%%%%%%%%%%%%%%
%%%%%%%%%%%%%%%%%%%%%%%%%%%%%%%%%No-region

%%%%%%%%%%%%%%%%%%%%%%%%%%%
\begin{figure}[htb]
\centerline{\includegraphics[width=1.0\textwidth]{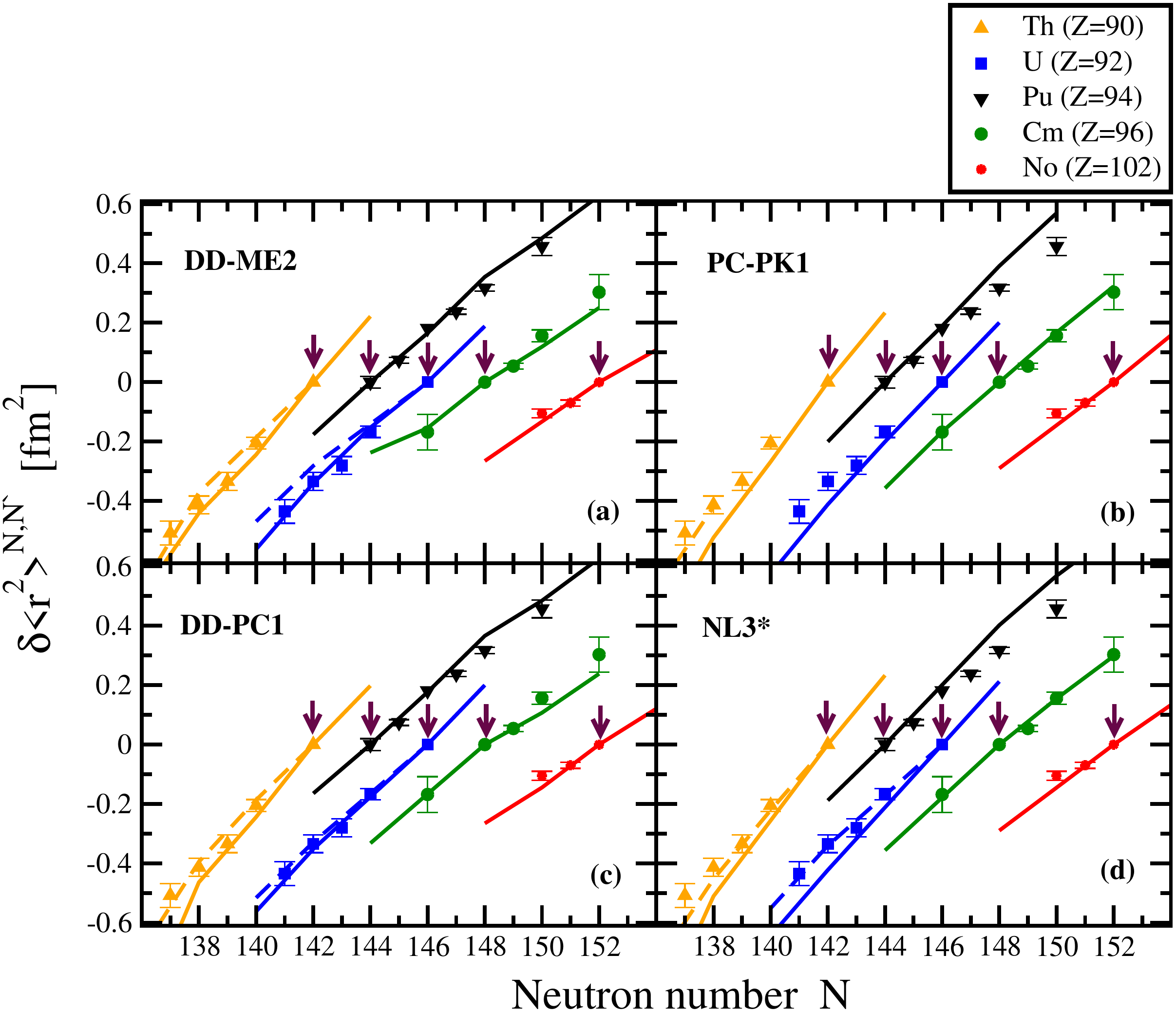}}
\caption{The experimental and calculated differential charge radii $\delta \left<r^2 \right>^{N,N'}$
of the indicated isotopic chains. The results of reflection symmetric and reflection asymmetric (with octupole
deformation) RHB calculations are shown by solid and dashed lines, respectively.  Note that the latter are
based on the results obtained in Ref.\ \cite{AAR.16} and are shown only for the nuclei which have non-zero
octupole deformation $\beta_3$.  The arrows indicate the reference nuclei in the isotopic
chains. The experimental data are taken from Refs.\ \cite{AM.13,Pu,No}.
}
\label{actinides+No}
\end{figure}
%%%%%%%%%%%%%%%%%%%%%%%%%%%%%%%%%No-region

  In the present paper, we focus on differential charge radii of deformed actinides
and light superheavy nuclei which were not covered in Ref.\ \cite{UAR.21}. The aim
is to understand the accuracy of the description of the $\delta \left<r^2 \right>^{N,N'}$
values in high-$Z$ systems and the impact of octupole deformation on differential charge
radii.  Note that due to the difficulties of the measurements of absolute charge radii in the
$Z>83$ nuclei,  the experimental data covers only differential charge radii [the only exception 
is uranium].  The experimental and calculated differential charge radii  of the Th, U, Pu, Cm,  and
No isotopic chains are compared in Fig. \ref{actinides+No}.  Note that only even-even nuclei are
considered in the calculations.  The format of the figure is similar to that of Fig. 2 of Ref.\ \cite{No}
in which the results of Skyrme DFT calculations with UNEDF1 and SV-min functionals are compared
with experimental data. Note that the RHB calculations indicate the presence of  octupole
deformation in low-$N$ Th and U isotopes (see Ref.\ \cite{AAR.16}) but, in a given nucleus,
it somewhat  depends on the employed CEDF and on the type of the DFT \cite{AAR.16,CAANO.20}.
The accounting of octupole deformation leads to an increase of the $\delta \left<r^2 \right>^{N,N'}$
value by approximately  $\approx 0.1$ fm$^2$ and leads to an improvement in the description
of differential charge radii in the Th and U isotopes.

   On average, the accuracy of the description of differential charge radii obtained with
covariant (DD-ME2, DD-PC1, PC-PK1, and NL3*) and non-relativistic  energy density functionals
(UNEDF1 and SV-min) is similar (compare Fig.\ \ref{actinides+No} in  the present paper with Fig. 2
in Ref.\ \cite{No}). Among  the employed CEDFs, the best agreement with experiment is obtained for the
Th isotopes by DD-ME2, DD-PC1 and NL3*, for the U isotopes
by DD-PC1 and NL3*, for the Pu isotopes by DD-ME2 and DD-PC1, and for the Cm isotopes by
PC-PK1 and NL3* (see Fig.\ \ref{actinides+No}). The results of the calculations for the No isotopes
are the same with all employed functionals\footnote{Note that the analysis of  the No isotopes 
performed in Ref.\ \cite{ADFAA.20}  suggests that the measurements of at least two atomic transitions 
would allow to disentangle the contributions of the changes in deformation and
charge radius into field isotopic shifts.}. The local deviations between theory and experiment
for $\delta \left<r^2 \right>^{N,N'}$ are most  likely due to the differences between experimental and
calculated energies of the deformed quasiparticle states (see Refs.\ \cite{AS.11,BQM.07,DABRS.15})
which leads to some dissimilarity of their contributions to charge radii. The same is true when the
results of the calculations with two functionals are compared (see Sect. IV of Ref.\ \cite{UAR.21} for a
discussion of the impact of the underlying single-particle structure and occupation pattern of the
single-particle states on differential charge radii of the spherical Pb nuclei).

%%%%%%%%%%%%%%%%%%%%%%%%%%%%%%%%%%%%%%%%
\section{New mechanisms leading to odd-even staggering in charge radii}
\label{sec-OES}
%%%%%%%%%%%%%%%%%%%%%%%%%%%%%%%%%%%%%%%%
 
   The description of odd-even staggering (OES)  in charge radii has been a
challenge for a long time.  In the most commonly accepted scenario, it is due to the gradient
terms both in the surface  part of the energy density functional and in the pairing
interaction (see Refs.\ \cite{FTTZ.94,FTTZ.00,RN.17}).  However,  this has recently been
challenged in Ref.\  \cite{OES-Pb-Hg.21} (see also Ref.\ \cite{OES-Pb-Hg.21-PRC} for a
more detailed study)  where it has been shown that the scattering of the occupation of different
single-particle states  between  even-even and odd nuclei in  the isotopic chain can lead to
OES in charge radii.  This scattering is due to the fact that particle-vibration coupling
(PVC)  leads to a change of the relative order of the single-particle states in odd-$N$
nuclei as compared with even-$N$ ones in which PVC plays a minor role.

\begin{figure}[htb]
	\centerline{\includegraphics[width=0.95\textwidth]{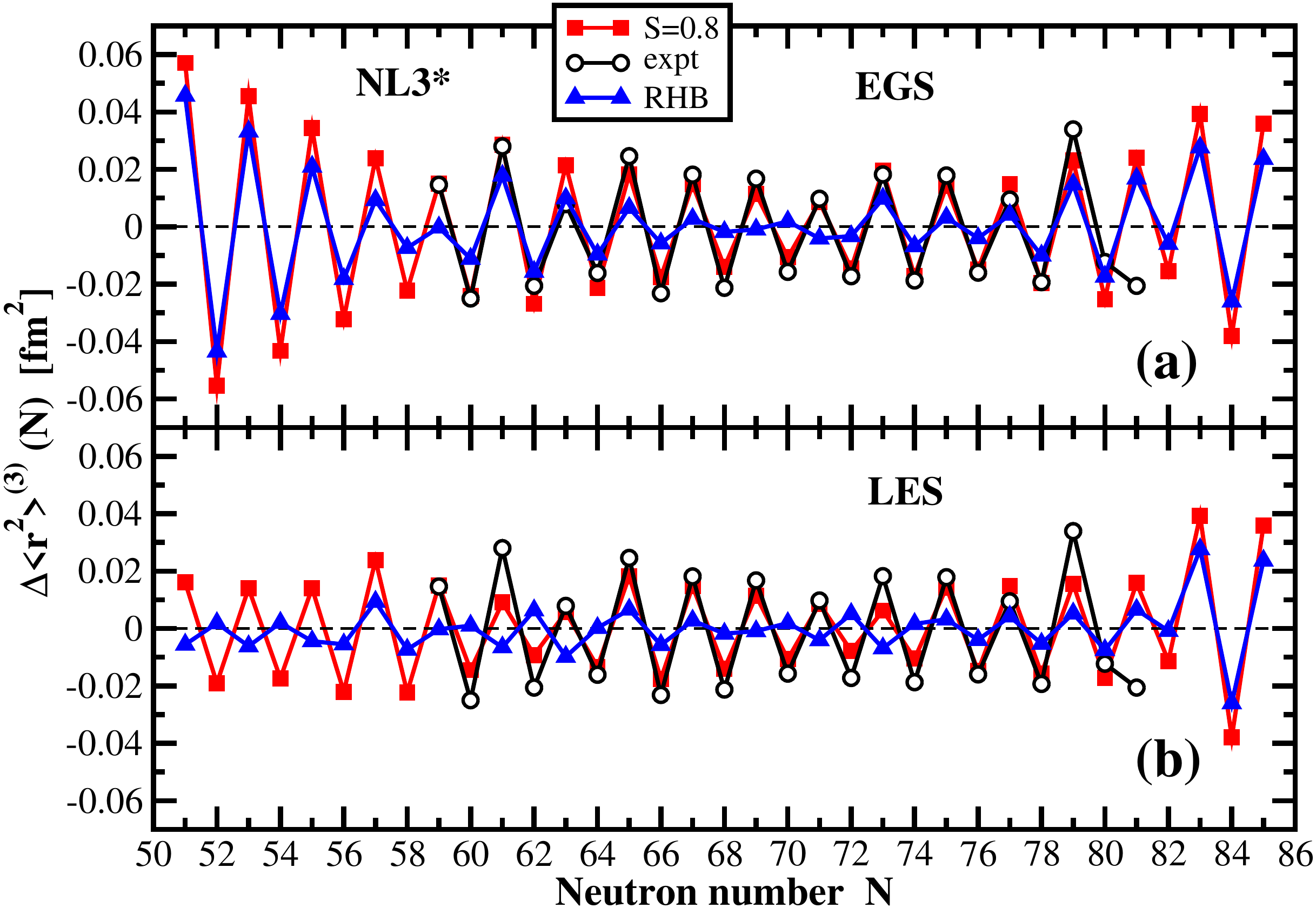}}
	\caption{OES in charge radii of the Sn isotopes. The experimental data are
	taken from Ref.\ \cite{AM.13}. The RHB results with the  NL3* CEDF are shown by blue lines with
	triangles for the LES and EGS procedures in odd-$N$ nuclei. The results  corrected
	for the fragmentation of the single-particle content of the dominant
         single-particle state in odd-$N$ nuclei  within the schematic model discussed in the
          text are shown by red lines with squares for  a representative  spectroscopic factor
	 $S = 0.8$.}
	\label{Sn-PES-PVC}
\end{figure}

  The analysis of Refs.\ \cite{OES-Pb-Hg.21,OES-Pb-Hg.21-PRC,UAR.21} also
clearly indicates the importance of taking into account both the structure of the ground
state in odd-$A$ nuclei and the fact that the DFT calculations do not reproduce the
structure of the ground states in odd-$A$ nuclei in more than half of the cases
(see Refs.\ \cite{BQM.07,LA.11,AS.11,AL.15}). These facts have been ignored
in the absolute majority of the studies of OES in charge radii. The importance of the
accounting of these facts is illustrated in Fig.\ \ref{Sn-PES-PVC} in which OES
in charge radii of the Sn isotopes obtained with two procedures (LES and EGS)
in the RHB calculations of odd-$N$ nuclei are compared. In the LES procedure,
the lowest in energy calculated configuration is used. In the EGS procedure, the
configuration with the spin and parity of the blocked state corresponding to those
of the experimental ground state is employed, although it is not necessarily the
lowest in energy.  The results of the RHB calculations with the LES procedure significantly
underestimate the magnitude of experimental OES and occasionally provide a
wrong phase of the OES (see Fig.\ \ref{Sn-PES-PVC}(b)). The use of the EGS
procedure significantly improves the description of both the phases and the
magnitude of OES (see Fig.\ \ref{Sn-PES-PVC}(a)). However, even in that case,
the magnitude of OES is underestimated on average by a factor of approximately
two.

%%%%%%%%%%%%%%%%%%%%%%%%%%%%%%%%%%%%%%%
%%%%%%%%%%%%%%%%%%%%%%%%%%%%%%%%%%%%%%%%

 The remaining discrepancies between theory and experiment suggest that the energy 
considerations discussed above are not sufficient and special attention has to be paid  
to the wavefunction of the ground state in odd-$N$ nuclei. There is a substantial difference 
between the ground states in even-even and odd-$A$ nuclei with stiff parabola-like potential 
energy surfaces (such as the Sn isotopes). Indeed, the impact of the correlations beyond 
mean-field on the wavefunction of the ground state is relatively small in even-even
nuclei, but it is substantial in odd-$A$ ones. The latter is due to the coupling of the single-particle
motion with phonons (particle-vibration coupling) which leads to a substantial fragmentation of
the wavefunction; this feature is confirmed in numerous experiments (see Refs.\  \cite{LR.06,LA.11,AL.15}
and references quoted therein).  This reduces the single-particle content  of the unpaired neutron
ground state in odd-$A$ nuclei to the value characterized by the spectroscopic factor $S$ ($S\leq 1.0$).
As a result, the pull provided by the odd neutron on the proton radii (see Refs.\ \cite{GSR.13,UAR.21} and
Sec.\ \ref{sec-self-const}) is reduced as compared with the mean-field value.  Assuming that at
the mean-field level the increase of charge radii on going from even to odd neutron number
in the even $Z$ isotopic chain is $\Delta (r_{ch})_{MF}^{qp} = r_{ch}(N+1) - r_{ch}(N)$ [here
$N$ is even], the accounting of the fragmentation of the  wavefunction in odd-$N$ nuclei
will reduce  this increase to $\Delta (r_{ch})_{frag} = S \Delta (r_{ch})_{MF}^{qp}$.  Using
this schematic model (see Ref.\ \cite{UAR.21} for more details), one can show  that the
fragmentation of the structure of the unpaired neutron in odd-$N$ nuclei due to 
particle-vibration coupling leads to an increase of the magnitudes of OES in charge radii and 
correct phases of the OES both in the LES and EGS  procedures for typical values of $S$ 
(see Fig.\ \ref{Sn-PES-PVC}). Both of these factors act in the direction of improving the 
agreement with experiment.

%%%%%%%%%%%%%%%%%%%%%%%%%%%%%%%%%%%%%%%%
\section{The role of self-consistency effects}
\label{sec-self-const}
%%%%%%%%%%%%%%%%%%%%%%%%%%%%%%%%%%%%%%%%

   As reviewed in the introduction of Ref.\ \cite{UAR.21},  the absolute majority of
the studies of differential  charge radii has been performed either in DFT or in ab initio
approaches.  However, it is well known that these types  of models have some deficiencies
in the description of spectroscopic properties related to the energies  of the single-particle states
and their wave functions  \cite{BQM.07,CSB.10,AS.11,DABRS.15,AL.15}. In contrast, the spherical
shell models with empirical interactions provide a better description of experimental spectroscopic
data in spherical nuclei  located in the vicinity  of doubly magic nuclei and  microscopic+macroscopic
(mic+mac) models based on phenomenological potentials such as the Woods-Saxon one do the
same in  the region of deformed nuclei.  However, we are not aware of successful investigations of
differential charge radii within these types  of models and naturally the question of the underlying
reasons emerges.

%%%%%%%%%%%%%%%%%%%%%%%%%%%%%%%%%%%%%%%
\begin{table}[htbp]
\caption{Proton single-particle rms radii $r_{p, i} = \sqrt{\left< r^2 \right>_{i}}$ [in $fm$] (columns 4 and 5)
and single-particle energies $e_i$ [in MeV] (columns 2 and 3)  of the indicated single-particle ($i$-th) % orbitals
in the nuclei $^{208}$Pb and $^{220}$Pb. The pairing is neglected in the calculations and they employ %the
NL3* functional. The column 6 shows the change of proton single-particle radii
    $\delta r_{p, i}  = r_{p, i} (^{220}{\rm{Pb}}) - r_{p, i} (^{208}\rm{Pb})$
        between the nuclei $^{208}$Pb and $^{220}$Pb. Note that the results for the single-particle states with
       principal quantum number $n=1$ are shown in bold.
} 
\begin{tabular}{|c|c|c|c|c|c|}
\hline
{subshell}  & $e_i$ ($^{208}$\rm{Pb}) & $e_i$ ($^{220}$\rm{Pb}) & $r_{p, i}$ ($^{208}$\rm{Pb}) &
 $r_{p, i}$ ($^{220}$\rm{Pb})& $\delta r_{p,i}$ \\
\hline
       1                     &             2           &             3           &                  4           &
       5                     &          6             \\ 
\hline
$1s_{1/2}$	& \bf{-48.886}	& \bf{-48.151}	& \bf{4.067759} & \bf{4.277595}  & \bf{0.209836}  \\
$1p_{3/2}$	& \bf{-43.193}	& \bf{-43.155}	& \bf{4.667454}	& \bf{4.889208}	  & \bf{0.221754}  \\
$1p_{1/2}$	& \bf{-42.510}	& \bf{-42.620}   & \bf{	4.575735}& \bf{	4.805227}	  & \bf{0.229492}  \\
$1d_{5/2}$	& \bf{-36.097}	& \bf{-36.853}	& \bf{5.11159}	& \bf{5.324115}	  & \bf{0.212525}  \\
$1d_{3/2}$	&\bf{-34.533}	& \bf{-35.636}	& \bf{4.972567}	& \bf{5.187897}	  & \bf{0.21533}    \\
$2s_{1/2}$	&  -30.844	        &	-32.215	& 4.458931	&	4.544538	  &  0.085607        \\
$1f_{7/2}$ 	& \bf {-28.042}	&\bf{	-29.573}	& \bf{5.487019}	& \bf{5.682079}	  & \bf{0.19506}    \\
$1f_{5/2}$ 	& \bf{-25.265}	&\bf{	-27.366}	& \bf{5.325364}	& \bf{5.512845}	  & \bf{0.187481} \\
$2p_{3/2}$	&	-20.890	&	-22.819	&  5.000532	&	5.032818	  &	0.032286    \\
$2p_{1/2}$	&	-19.839	&	-21.843	&  5.004416	&	5.03297	  &	0.028554    \\
$1g_{9/2}$	& \bf{-19.369}	&\bf{	-21.587}   & \bf{5.82501}	& \bf{5.998839}  & \bf{0.173829} \\
$1g_{7/2}$	& \bf{-15.175}   &\bf{	-18.149}   & \bf{	5.673305}& \bf{	5.82303}	  & \bf{0.149725} \\
$2d_{5/2}$	&	-11.154	&	-13.413	&	5.53793   &	5.542713	  &	0.004783    \\
$1h_{11/2}$	& \bf{-10.335}   &\bf{	-13.120}	&\bf{	6.14104}	& \bf{6.291571}	  & \bf{0.150531} \\
$2d_{3/2}$	&	-9.537	&	-11.909	&	5.572499	& 5.577024	  &	0.004525    \\
$3s_{1/2}$	&	-8.428	&	-10.725	&5.495602	& 5.482959	  &	-0.012643   \\
\hline
\hline
\end{tabular}
\label{table-sp-radii}
\end{table}
%%%%%%%%%%%%%%%%%%%%%%%%%%%%%%%%%%%%%%%%%%%%%%%%%

In order to answer this question, we consider single-particle rms radii of the
proton orbitals occupied in $^{208}$Pb and $^{220}$Pb as obtained in the calculations without
pairing (see Table 1).  These nuclei differ by 12 neutrons which completely occupy the
$\nu 1i_{11/2}$ subshell\footnote{We consider a full neutron subshell in order to magnify the
impact of the occupation of neutron states on the proton single-particle rms radii.} (see discussion in
Ref.\ \cite{UAR.21}). One can see\footnote{Similar impact of the occupation of the $1i_{11/2}$ neutron 
subshell on charge radii of proton subshells is seen also in the Skyrme DFT calculations of Ref.\ 
\cite{GSR.13}.} that the pull provided by neutrons in this subshell
on rms radii of proton single-particle states depends on their principal quantum number 
$n$.  This pull is
quantified by $\delta r_{p,i}$ (see Table 1) and it is the largest for the $n=1$ orbitals (especially
those located at the bottom of the potential). It is reduced drastically for the $n=2$ and 3 proton
orbitals. Note that the proton rms point radius entering into Eq.\ (\ref{r_charge}) is given by
\begin{equation}
\left< r^2 \right>_p = \sum_i (2j_i+1) r_{p,i}^2
\end{equation}
where the sum runs over all occupied spherical subshells and  {\bf $(2j_i+1)$ } is the degeneracy of the $i$-th
subshell.  As a result, the differential charge radius between two nuclei (see Eq.\ (\ref{diff-radii})) can
be rewritten as
\begin{equation}
\delta \left< r^2 \right>^{N,N'}_p = \sum_i (2j_i+1) (r_{p,i}^2(N) - r_{p,i}^2(N'))
\end{equation}
One can see that it depends on the pull on the rms radii of all occupied  proton subshells provided
by the occupation of neutron states.  Thus the introduction of the core (as it is done in the spherical
shell model) leads to a neglect of the pull provided by the extra neutron(s) on the proton single-particle
states forming the core.  This introduces uncontrollable errors in the calculations of differential charge 
radii and thus severely limits the applicability of spherical shell model to the description of this 
observable.
  
    In a similar fashion,  the lack of self-consistency effects is expected to affect the description
of differential charge radii in the mic+mac model. This is because the addition of neutron does not affect 
the proton subsystem in a self-consistent manner on the level of single-particle subshells (as in Table 
\ref{table-sp-radii}).  For example, in the Woods-Saxon potential it affect the  total radius of nucleus $R$ 
only via mass dependence $R=1.2 A^{1/3}$: this means that the occupation  of the neutron $1i_{11/2}$ 
and $2g_{9/2}$ subshells in the $N>126$ Pb isotopes will lead to the same differential charge radii contrary 
to the results of self-consistent calculations (see Refs.\ \cite{SLR.93,RF.95,OES-Pb-Hg.21,OES-Pb-Hg.21-PRC}.) 
Moreover, there is a lack of self-consistency in the definition  of the radial properties of the density 
distributions in the macroscopic (liquid drop) and microscopic  (single-particle potential) parts of the 
mic+mac model.  To our knowledge, this aspect of the problem has not been studied in details.  However, 
the physical observables similar to  $\delta \left < r^2 \right>_p^{N,N'}$, namely, relative charge quadrupole 
moments of superdeformed  bands are affected by the lack of self-consistency between microscopic and 
macroscopic parts \cite{KRA.98}.

%%%%%%%%%%%%%%%%%%%%%%%%%%%%%%%%%%%%%%%%
\section{Conclusions}
\label{conclusions}
%%%%%%%%%%%%%%%%%%%%%%%%%%%%%%%%%%%%%%%%

 Recent applications of covariant density functional theory to the analysis and description
of charge radii and related indicators have been reviewed. It covers global accuracy of the
description of experimental absolute and differential charge radii and the evaluation of 
related theoretical statistical and systematic uncertainties. This theory provides  reasonably 
good description of these physical observables already at the mean field level. 
The discrepancies between theory and experiment have two sources: the deficiencies
in the description of the energies  of the single-particle states and the neglect of the
correlations beyond mean field.  For example, the former affects the reproduction of the 
evolution of differential charge radii in isotopic chains which cross the $N=28$ and $N=82$ 
neutron shell closures. The later shows itself in two areas: (i) in the description of charge 
radii in shape coexistent nuclei or in the nuclei with soft potential energy surfaces [accurate
description of such systems requires the methods such as generator coordinate method or 
5-dimensional collective Hamiltonian] and  (ii) in the description of odd-even staggering in 
charge radii since the energies and the wave functions of one-quasiparticle states in odd-$A$ 
nuclei are affected [strongly in spherical nuclei] by particle-vibration coupling.

   In addition, new calculations reveal that differential charge radii in deformed actinides and 
light superheavy nuclei are well described in the CDFT calculations. However, it is important 
to take into account octupole deformation in low-$N$ Th and U nuclei.  Finally, we analyzed 
the importance of self-consistency effects on the accurate description of differential charge.
This analysis reveals the challenging situation. The CDFT as well as the Skyrme and Gogny 
DFTs 
%properly 
take  these effects into account on the level of single-particle states but
suffer from some deficiencies in the description of their energies. In contrast, the models
which are specifically fitted for a very accurate description of the single-particle energies 
suffer from the neglect of these self-consistency effects either due to introduction of 
the closed core (as in the most of spherical shell models) or due to the split of the model 
space into macroscopic and microscopic part (as in the mic+mac models).

%%%%%%%%%%%%%%%%%%%
\section*{Acknowledgements}
%%%%%%%%%%%%%%%%%%%

This material is based upon work supported by the U.S. Department of Energy,
Office of Science, Office of Nuclear Physics under Award No. DE-SC0013037.
PR acknowledges partial support from the Deutsche Forschungsgemeinschaft
(DFG, German Research Foundation) under  Germany Excellence
Strategy EXC-2094-390783311, ORIGINS.

%\section*{References}


\begin{thebibliography}{99}

\bibitem{AM.13} I. Angeli and K. P. Marinova,
%Table of experimental
%nuclear ground state charge radii: An update,
At. Data Nucl. Data Tables 99, 69 (2013).

\bibitem{CMP.16} P. Campbell, I. Moore, and M. Pearson,
% Laser spectroscopy for nuclear structure physics,
Prog. Part. Nucl. Phys. 86, 127 (2016).

\bibitem{VALR.05} D. Vretenar, A. V. Afanasjev, G. A. Lalazissis, and P. Ring,
                              Phys. Rep. 409, 101 (2005).

\bibitem{SLR.93} M. M. Sharma, G. A. Lalazissis, and P. Ring,
                            Phys. Lett. B 317, 9 (1993).

\bibitem{RF.95} P.-G. Reinhard and H. Flocard,
%Nuclear effective forces and isotope shifts,
Nucl. Phys. A 584, 467 (1995).

\bibitem{TMR.09} Y. Tian, Z. Y. Ma, and P. Ring,
%A  finite range pairing force for density functional theory in
%superfluid nuclei,
Phys. Lett. B 676, 44 (2009).

\bibitem{AARR.14} S.~ E.~ Agbemava and A.~ V.~ Afanasjev and D.~ Ray and P.~ Ring,
% TITLE={Global performance of covariant energy density
%        functionals: Ground state observables of even-even
%        nuclei and the estimate of theoretical uncertainties},
  Phys. Rev. C 89, 054320 (2014).

\bibitem{NL3*} G.~ A.~ Lalazissis, S.~ Karatzikos, R.~ Fossion, 
        D.~ Pe{\~n}a~Arteaga, A.~ V.~ Afanasjev and P.~ Ring,
        Phys. Lett. B671, 36 (2009).
 
\bibitem{DD-ME2} G.~ A.~ Lalazissis, T.~ Nik{\v{s}}i{\'{c}}, D.~ Vretenar 
                             and P.~ Ring, Phys. Rev. C 71, 024312 (2005).

\bibitem{DD-MEdelta}  X.\ Roca-Maza, X.\ Vi{\~n}as, M.\ Centelles,
                                    P.\ Ring and P.\ Schuck,
                                    Phys. Rev. C 84, 055309 (2011).

 \bibitem{DD-PC1} T.~ Nik\v{s}i\'{c}, D.~ Vretenar and P.~ Ring,
             Phys. Rev. C 78, 034318 (2008).

\bibitem{PC-PK1}  P.~ W.~ Zhao, Z.~ P.~ Li, J.~ M.~ Yao and J.~ Meng,
                             Phys. Rev. C 82, 054319 (2010).  
 
\bibitem{AAT.19} S. E. Agbemava, A. V. Afanasjev, and A. Taninah,
Phys. Rev. C 99, 014318 (2019).

\bibitem{AA.16}  A. V. Afanasjev and S. E. Agbemava, Phys. Rev.
C 93, 054310 (2016).


\bibitem{UAR.21} U. C. Perera, A. V. Afanasjev, and P. Ring, subm. to
Phys. Rev. C (2021), see also nuclear theory arXiv:2108.02245

\bibitem{FP.75} S. Frauendorf and V. V. Pashkevich,
% On oblate-prolate  transition in the ground state rotational band of light
% mercury isotopes,
Phys. Lett. B 55, 365 (1975).

\bibitem{DGLGHPPB.10} J.-P. Delaroche, M. Girod, J. Libert, H. Goutte, S. Hilaire,
S. Peru, N. Pillet, and G. F. Bertsch,
% Structure of even-even nuclei using a mapped collective hamiltonian
% and the d1s gogny interaction,
Phys. Rev. C 81, 014303 (2010).

\bibitem{Otten} E. W. Otten, Nuclear radii and moments of unstable
isotopes, in Treatise on Heavy Ion Science: Volume 8:
Nuclei Far From Stability, edited by D. A. Bromley
(Springer, Boston, MA, (1989), pp. 517638.

\bibitem{Hg-mid}  S. Sels {\it et al}, Phys. Rev. C 99, 044306 (2019).

\bibitem{BHR.03} M. Bender, P.-H. Heenen, and P.-G. Reinhard, Rev. Mod. Phys.
                            75, 121 (2003).

\bibitem{BB.77} J. Boguta and R. Bodmer, Nucl. Phys. A 292, 413 (1977).

\bibitem{DNR.14} J. Dobaczewski, W. Nazarewicz, and P.-G. Reinhard, J. Phys. G
                             41, 074001 (2014).

\bibitem{GDKTT.13} Y. Gao, J. Dobaczewski, M. Kortelainen, J. Toivanen, and
                                 D. Tarpanov, Phys. Rev. C 87, 034324 (2013).

\bibitem{OES-Pb-Hg.21-PRC} T. D. Goodacre {\it et al}, Phys. Rev. C, in press

\bibitem{GV.20} R. F. Garcia-Ruiz and A. Vernon, Eur. Phys. J. A 56,
                         136 (2020).

\bibitem{AAR.16} S.~ E.~ Agbemava and A.~ V.~ Afanasjev and P.~ Ring,
% TITLE={Octupole deformation in the ground states of
%        even-even nuclei: a global analysis within the
%        covariant density functional theory},
Phys. Rev. C 93, 044303 (2016).

\bibitem{Pu} A. Voss, V. Sonnenschein, P. Campbell, B. Cheal, T. Kron, I. D. Moore,
                    I. Pohjalainen, S. Raeder, N. Trautmann, and K. Wendt,
%{Nuclear energy density optimization: Large deformations,}
                                        Phys. Rev. A 95, 032506 (2017).                    

\bibitem{No} S. Raeder et al.,
%{Probing Sizes and Shapes of Nobelium Isotopes by Laser Spectroscopy,}
Phys. Rev. Lett. 120, 232503 (2018).

\bibitem{ADFAA.20} Saleh O. Allehabi, V. A. Dzuba, V. V. Flambaum, 
                                 A. V. Afanasjev and S. E. Agbemava, Phys. Rev. C 102, 024326 (2020).

\bibitem{CAANO.20} Yunchen Cao, S.E. Agbemava, A.V. Afanasjev, W. Nazarewicz
                                 and E. Olsen, Phys. Rev. C 102, 024311 (2020).

\bibitem{AS.11}  A.~ V.~ Afanasjev and S.~ Shawaqfeh, Phys. Lett. B 706, 177 (2011).
% TITLE={Deformed  one-quasiparticle states in covariant
%        density functional theory},

\bibitem{BQM.07} L.~ Bonneau, P.~ Quentin and P.~ M{\"o}ller,
             Phys. Rev. C 76, 024320 (2007).

\bibitem{DABRS.15}  J. Dobaczewski, A. V. Afanasjev, M. Bender, L. M. Robledo
                                 and Yue Shi, Nucl. Phys. A 944, 388 (2015).
%title = "Properties of nuclei in the nobelium region studied within
%       the covariant, Skyrme, and Gogny energy density functionals ",

\bibitem{FTTZ.94} S. A. Fayans, S. V. Tolokonnikov, E. L. Trykov, and D. Zawischa,
% Isotope shifts within the energy-density functional approach with density
% dependent pairing,
  Phys. Lett. B 338, 1994 (1).

\bibitem{FTTZ.00} S. A. Fayans, S. V. Tolokonnikov, E. L. Trykov, and D. Zawischa,
% Nuclear isotope shifts within the local energy-density functional approach,
Nucl. Phys. A 676, 49 (2000).

\bibitem{RN.17} P.-G. Reinhard and W. Nazarewicz,
                           Phys. Rev. C 95, 064328 (2017).

\bibitem{OES-Pb-Hg.21} T. D. Goodacre {\it et al}, Phys. Rev. Let. 126, 032502
                                        (2021).

\bibitem{LA.11} E. V. Litvinova and A. V. Afanasjev, Phys. Rev. C 84, 014305 (2011).

\bibitem{AL.15} A. V. Afanasjev and E. Litvinova,
%{Impact of collective
%	vibrations on quasiparticle states of open-shell odd-mass
%	nuclei and possible interference with the tensor force,}
Phys. Rev. C 92, 044317 (2015).

\bibitem{LR.06} E. Litvinova and P. Ring, Phys. Rev. C 73, 044328 (2006).

\bibitem{GSR.13} P. M. Goddard, P. D. Stevensson, and A. Rios,  Phys.
              Rev. Lett. 110, 032503 (2013).

\bibitem{CSB.10} G. Col{\'o}, H. Sagawa, and P. F. Bortignon,
%Effect of particle-vibration coupling on single-particle states: A consistent study 
%within the skyrme framework,}
Phys. Rev. C 82, 064307 (2010).

\bibitem{KRA.98} L.~B.~ Karlsson,  I. Ragnarsson and S. {\AA}berg,
                             Nucl. Phys. A 639, 654 (1998).
                             %Polarization effects in superdeformed nuclei



\end{thebibliography}
\end{document}